\documentclass[12pt]{article}

\usepackage{epsf}

\renewcommand{\vec}[1]{{\mathbf{#1}}}

\newcommand{\deltwo}{\delta^{(2)}}

\newcommand{\half}{{\textstyle \frac{1}{2}}}

\begin{document}

\begin{flushright}
  hep-ph/0205208
\end{flushright}

\vspace{\baselineskip}

\begin{center}
\textbf{\Large Generalized parton distributions \\[0.3em]
in impact parameter space} \\ 
\vspace{4\baselineskip}
{\large M. Diehl} \\
\vspace{1\baselineskip}
\textit{Institut f\"ur Theoretische Physik E, RWTH Aachen, 52056
Aachen, Germany} \\
\vspace{5\baselineskip}
\textbf{Abstract}\\
\vspace{1\baselineskip}
\parbox{0.9\textwidth}{We study generalized parton distributions in
the impact parameter representation, including the case of nonzero
skewness $\xi$.  Using Lorentz invariance, and expressing parton
distributions in terms of impact parameter dependent wave functions,
we investigate in which way they simultaneously describe longitudinal
and transverse structure of a fast moving hadron.  We compare this
information with the one in elastic form factors, in ordinary and in
$k_T$-dependent parton distributions.}
\end{center}
\vspace{\baselineskip}
%
%


\section{Introduction}

In recent years, generalized parton distributions (GPDs)
\cite{Muller:1994fv,Ji:1997ek,Radyushkin:1997ki} have become a subject
of considerable theoretical and experimental activity.  Much of the
interest in these quantities has been triggered by their potential to
help unravel the spin structure of the nucleon, as they contain
information not only on the helicity carried by partons, but also on
their orbital angular momentum~\cite{Ji:1997ek}.  This is, however,
only one of the aspects where GPDs can provide qualitatively new
insights on how a hadron looks like at the level of quarks and gluons.
A look at the variables on which GPDs depend immediately reveals that
they simultaneously carry information on both the longitudinal and the
transverse distribution of partons in a fast moving hadron.  The
physical picture becomes particularly intuitive after a Fourier
transform from transverse momentum transfer to impact
parameter~\cite{Burkardt:2000za}.  Analogies with other imaging
techniques have been highlighted in~\cite{Ralston:2001xs}: the Fourier
transform also occurs in geometrical optics, with light rays
corresponding to momentum space and an image plane to transverse
position, or in $X$-ray diffraction of crystals, where the diffraction
pattern has to be Fourier transformed to recover spacial information.

Burkardt has shown that for vanishing transfer of longitudinal
momentum (i.e.\ $\xi=0$ in standard notation) GPDs transformed to
impact parameter space have the interpretation of a density of partons
with longitudinal momentum fraction $x$ and transverse distance $b$
from the proton's center~\cite{Burkardt:2000za}.  It is natural to ask
how this situation looks like at nonzero $\xi$, which is relevant for
most processes where GPDs can be accessed.  This is the purpose of the
present work.  

In section~\ref{sec:lorentz} we will see how the interplay between
longitudinal and transverse degrees of freedom is dictated by Lorentz
symmetry.  We shall then study an explicit realization of this
interplay in section~\ref{sec:overlap}: using the technique
of~\cite{Diehl:2000xz} we will write GPDs in terms of the light-cone
wave functions for the target, in the mixed representation of
longitudinal momentum and transverse distance of the partons.  We
discuss some special cases of this representation in
section~\ref{sec:special}, and the question of resolution scales in
section~\ref{sec:scales}.  Apart from GPDs, transverse information on
partons in a hadron is also contained in parton distributions that
depend on the tranverse momentum of the struck parton.  We will
compare the different nature of the information contained in these
quantities in section~\ref{sec:unintegrated}, before summarizing our
main results in section~\ref{sec:sum}.  For definiteness we will
present our discussion for quark GPDs in a nucleon here; its
generalization to gluons and targets with different spin is
straightforward.


\section{Lorentz invariance and its consequence}
\label{sec:lorentz}

\subsection{From momentum transfer to impact parameter}
\label{sec:impact}

To describe the kinematics of GPDs we will represent a four-vector $v$
in terms of light-cone coordinates $v^\pm = (v^0 \pm v^3)/\sqrt{2}$
and its transverse components $\vec{v} = (v^1,v^2)$.  To avoid
proliferation of subscripts like ``$\perp$'' we will reserve bold-face
for two-dimensional transverse vectors throughout.  The momenta and
helicities of the initial (final) nucleon are $p$ ($p'$) and $\lambda$
($\lambda'$).  We use Ji's variables $x$ and $\xi$ for plus-momentum
fractions relative to the average momentum $P = \frac{1}{2} (p+p')$,
in particular we have $(p-p')^+ = \xi\, (p+p')^+$.  In the light-cone
gauge $A^+ =0$, the matrix elements defining GPDs are
\begin{eqnarray} 
  \label{quark-unpol-def}
{\cal H}_{\lambda'\lambda} &=&
\frac{1}{\sqrt{1-\xi^2}} \,
\int\frac{d z^-}{4\pi}\; e^{i x P^{+}z^-}
\langle p^{\,\prime},\lambda'|
   \,\bar{q}(0,-\half z^-,\vec{0})\,\gamma^+ 
         q(0,\half z^-,\vec{0})\, |p,\lambda\rangle
\\[0.5em]
&=&  \frac{1}{2 P^{+}\sqrt{1-\xi^2}}\, \bar u(p^{\,\prime},\lambda')
 \left[ \, H(x,\xi,t)\, \gamma^+
         + E(x,\xi,t)\, \frac{i\sigma^{+\alpha}(p'-p)_\alpha}{2m} 
 \, \right] u(p,\lambda)
\nonumber 
\end{eqnarray} 
for the sum over quark helicities, and 
\begin{eqnarray} 
  \label{quark-pol-def}
\widetilde{\cal H}_{\lambda'\lambda} &=&
\frac{1}{\sqrt{1-\xi^2}} \,
\int\frac{d z^-}{4\pi}\; e^{i x P^{+}z^-}
\langle p^{\,\prime},\lambda'|
   \,\bar{q}(0,-\half z^-,\vec{0})\,\gamma^+ \gamma_5 \,
         q(0,\half z^-,\vec{0})\, |p,\lambda\rangle
\\[0.5em]
&=&  \frac{1}{2 P^{+}\sqrt{1-\xi^2}}\, \bar u(p^{\,\prime},\lambda')
 \left[ \, \tilde{H}(x,\xi,t)\, \gamma^+\gamma_5
         + \tilde{E}(x,\xi,t)\, \frac{(p'-p)^+ \gamma_5}{2m}
 \, \right] u(p,\lambda)
\nonumber 
\end{eqnarray} 
for the helicity difference, where $m$ denotes the proton mass and
position arguments for fields are given in the form $q(z) = q(z^+,
z^-, \vec{z})$.  Note that these definitions do \emph{not} refer to
any particular choice of transverse momenta $\vec{p}$ and
$\vec{p}'$\,: Lorentz invariance ensures that $H$, $E$, $\tilde{H}$,
$\tilde{E}$ only depend on $x$, $\xi$ and the invariant momentum
transfer
\begin{equation}
  \label{eq:t}
-t = -(p'-p)^2 = \frac{4 \xi^2 m^2}{1-\xi^2} 
                 + (1-\xi^2)\, \vec{D}^2
\end{equation}
with
\begin{equation}
  \label{eq:D}
  \vec{D} = \frac{\vec{p}'}{1-\xi} - \frac{\vec{p}}{1+\xi} .
\end{equation}
That $t$ depends on $\vec{p}$ and $\vec{p}'$ only through their
combination in (\ref{eq:D}) is a consequence of its invariance under
transverse boosts~\cite{Kogut:1970xa,Brodsky:1989pv}.  These are
Lorentz transformations which transform a four-vector $k$ according to
\begin{equation}
 \label{transv-boost}
k^+ \to k^+ , \qquad \vec{k} \to \vec{k} - k^+ \vec{v} ,
\end{equation}
and $k^-$ such that $k^2$ stays invariant.  Here $\vec{v}$ is a
transverse vector parameterizing the transformation.  Since $p'^+ =
(1-\xi) P^+$ and $p^+ = (1+\xi) P^+$, both vectors on the right-hand
side of (\ref{eq:D}) are shifted by the same amount $P^+ \vec{v}$, and
their difference remains the same.

In order to have simple properties of the matrix elements
(\ref{quark-unpol-def}) and (\ref{quark-pol-def}) under these
transformations, it is appropriate to choose the polarization states
of the protons such that they have definite light-cone
helicities~\cite{Kogut:1970xa,Brodsky:1989pv}.  The state vector and
spinor for a proton with momentum components $(p^+,\vec{p})$ is then
obtained from the one with $(p^+,\vec{0})$ by a transverse boost
(\ref{transv-boost}).\footnote{In contrast, the corresponding state
with definite (usual) helicity is obtained from the one with
$(p^+,\vec{0})$ by a spatial rotation.}
The matrix elements ${\cal H}_{\lambda'\lambda}$ and $\widetilde{\cal
H}_{\lambda'\lambda}$ are hence invariant under transverse boosts, and
depend on the proton momenta only through the variables $\xi$ and
$\vec{D}$, as do the GPDs $H$, $E$, $\tilde{H}$, $\tilde{E}$
themselves.  With the phase convention for proton spinors given
in~\cite{Diehl:2001pm} we have
\begin{eqnarray}
  \label{hel-unpol}
{\cal H}_{++} &=& {\cal H}_{--} \;=\; 
           H - \frac{\xi^2}{1-\xi^2}\, E \,,
\nonumber \\
{\cal H}_{-+} &=& - ({\cal H}_{+-})^* \;=\; 
           \frac{D^1 + i D^2}{2m}\; E \,,
\end{eqnarray}
and
\begin{eqnarray}
  \label{hel-pol}
\widetilde{\cal H}_{++} &=& - \widetilde{\cal H}_{--} \;=\; 
           \tilde{H} - \frac{\xi^2}{1-\xi^2}\, \tilde{E} \,,
\nonumber \\
\widetilde{\cal H}_{-+} &=& (\widetilde{\cal H}_{+-})^* \;=\; 
           \frac{D^1 + i D^2}{2m}\; \xi \tilde{E} \,.
\end{eqnarray}
We can now transform these matrix elements to impact parameter space,
\begin{eqnarray}
{\cal I}_{++}(x,\xi,\vec{b}) &=& 
\int \frac{d^2\vec{D}}{(2\pi)^2}\; e^{-i\, \vec{D}\cdot \vec{b}}\, 
                    {\cal H}_{++}(x,\xi,\vec{D})
\nonumber \\
 &=& \frac{1}{4\pi} \int_0^\infty \! d (\vec{D}^2)\,
         J_0\Big( |\vec{D}| |\vec{b}| \Big) \left( H 
              - \frac{\xi^2}{1-\xi^2}\, E \right) ,
\nonumber \\
{\cal I}_{-+}(x,\xi,\vec{b}) &=& 
\int \frac{d^2\vec{D}}{(2\pi)^2}\; e^{-i\, \vec{D}\cdot \vec{b}}\,
                    {\cal H}_{-+}(x,\xi,\vec{D})
\nonumber \\
 &=& \frac{1}{4\pi}\, \frac{b^2 - i b^1}{|\vec{b}|}
         \int_0^\infty \! d (\vec{D}^2)\,
         J_1\Big( |\vec{D}| |\vec{b}| \Big)\; 
              \frac{|\vec{D}|}{2m}\, E \,,
  \label{I-def}
\end{eqnarray}
with analogous relations for the other matrix elements.  Notice that
$E$ and $\tilde{E}$ parameterize both proton helicity conserving and
nonconserving transitions and thus appear twice, weighted by different
Bessel functions $J_0$ and $J_1$.

We remark that there is some freedom of choice in introducing impact
parameter GPDs.  Instead of defining the Fourier transform with
respect to $\vec{D}$, one could for instance use the vector
$(1-\xi^2)\, \vec{D}$ (which coincides with $\vec{p}'-\vec{p}$ in a
frame where $\vec{p}'= -\vec{p}$) or the vector $(1-\xi)\, \vec{D}$
(which is equal to $\vec{p}'$ when $\vec{p}=0$).  The corresponding
GPDs are related to those in (\ref{I-def}) by an overall factor and a
rescaling of the impact parameter.  Just as the different ways to
parameterize the plus-momentum fractions of the partons
\cite{Ji:1997ek,Radyushkin:1997ki}, such a change of variables does
not change the physics content of the GPDs.


\subsection{Probing transversely localized protons}
\label{sec:probing}

To discuss the spatial structure of a proton in the transverse plane
we now introduce proton states which are transversely localized.  To
avoid infinities in intermediate steps we use wave packets
\begin{equation}
\int \frac{d^2\vec{p}\, dp^+}{16\pi^3 p^+}\,  
     \Phi(p) |p,\lambda\rangle ,
  \label{wave-packet}
\end{equation}
which we choose to have definite plus-momentum, $\Phi(p) = p^+\,
\delta(p^+ - p_0^+)\, \Phi_\perp(\vec{p})$.  We find it convenient to
use a Gaussian shape for $\Phi_\perp(\vec{p})$, which will allow us to
perform integrations explicitly.  Abbreviating
\begin{equation}
G(\vec{x},\sigma^2) = 
  \exp\left( - \frac{\vec{x}^2}{2\sigma^2} \right)
\end{equation}
we thus define states centered around $\vec{b}_0$ with an accuracy of
order $\sigma$,
\begin{eqnarray}
 \label{proton-smeared}
|p^+, \vec{b}_0, \lambda\rangle_{\sigma} &=& 
   \int \frac{d^2\vec{p}}{16\pi^3}\, 
     e^{-i \vec{p}\cdot\vec{b}_0}\, 
     G\!\left(\vec{p}, \frac{1}{\sigma^2} \right)\, 
   |p^+, \vec{p}, \lambda\rangle
\\
&=& \int d^2\vec{b}\, \frac{1}{2\pi\sigma^2}\, 
     G(\vec{b}-\vec{b}_0, \sigma^2)\, |p^+, \vec{b}, \lambda\rangle ,
\nonumber 
\end{eqnarray}
where 
\begin{equation}
|p^+, \vec{b}, \lambda\rangle 
  = \int \frac{d^2\vec{p}}{16\pi^3}\, e^{-i \vec{p}\cdot\vec{b}}\,
    |p^+, \vec{p}, \lambda\rangle
  = \lim_{\sigma\to 0} |p^+, \vec{b}, \lambda\rangle_{\sigma}
 \label{proton-sharp}
\end{equation}
is completely localized in the transverse plane.  Of course the proton
is an extended object, and we will see in section~\ref{sec:overlap}
that more precisely it is the ``transverse center of momentum'' which
is localized for $|p^+, \vec{b}, \lambda\rangle$ and smeared out by an
amount $\sigma$ for $|p^+, \vec{b}, \lambda\rangle_{\sigma}$.  The
normalization of these states is
\begin{eqnarray}
{}_{\sigma} \langle p'^+, \vec{b}', \lambda' \,| \, 
     p^+, \vec{b}, \lambda\rangle_{\sigma} 
&=& p^+\, \delta(p'^+ - p^+)\,
       \frac{1}{16\pi^2\sigma^2}\, G(\vec{b}'-\vec{b}, 2\sigma^2)\,
    \delta_{\lambda'\lambda}
  \label{proton-norm}
\\[0.4em]
& \stackrel{\sigma\to 0}{=} & 
    p^+\, \delta(p'^+ - p^+)\,
       \frac{1}{4\pi}\, \deltwo(\vec{b}'-\vec{b})\,
     \delta_{\lambda'\lambda} .
\nonumber 
\end{eqnarray}
 
Notice that the relativistically invariant integration element in
(\ref{wave-packet}) does not mix $\vec{p}$ and $p^+$.  This is not the
case if the three-momentum components are taken as independent
variables, since in $d^2\vec{p}\, d p^3 / (2p^0)$ with $p^0 =
(\vec{p}^2 + (p^3)^2 + m^2)^{1/2}$ the dependence on $\vec{p}$ and
$p^3$ only decouples in limiting cases.  Forming wave packets with
definite $p^+$ instead of definite $p^3$ thus keeps our expressions
simple.  Notice however that for fixed $p^+$ the momentum component
$p^3 = (p^+ - p^-) /\sqrt{2}$ becomes negative for very large
$|\vec{p}|$ since $p^- = (\vec{p}^2 + m^2) /(2p^+)$.  If we want to
have the interpretation of a proton moving with a well-defined
longitudinal momentum, we must restrict the relevant range of
integration in (\ref{proton-smeared}) to $|\vec{p}| \ll p^+$ and thus
can only have a transverse localization $\sigma \gg 1/ p^+$, as
observed in~\cite{Burkardt:2000za}.  We should thus choose a frame
with large $p^+$.  This is not a restriction for our purposes, since
it is just in such a frame where we have the natural interpretation of
the proton as a bunch of partons, probed in a process where the GPD is
measured.  We also remark that if $p^+ \gg m$ and $p^+ \gg |\vec{p}|$,
the light-cone helicity states we are using here coincide with usual
helicity states up to corrections of order $m /p^+$,
see~\cite{Diehl:2001pm}.

We now want to take the matrix elements of the  operator
\begin{equation}
{\cal O}_{\bar{q}q}(\vec{z}) = 
   \int\frac{d z^-}{4\pi}\; e^{i x P^{+}z^-} \,
     \bar{q}(0,-\half z^-,\vec{z})\,\gamma^+ 
             q(0,\half z^-,\vec{z}) ,
\end{equation}
which defines the GPDs $H$ and $E$, between transversely localized
proton states.  The discussion for $\tilde H$ and $\tilde E$ is
completely analogous.  Using (\ref{quark-unpol-def}) and
(\ref{proton-smeared}) we get
\begin{eqnarray}
\lefteqn{
\rule[-1.3ex]{0pt}{1ex}_{\sigma} 
  \Big\langle p'^+, \vec{b}'_0 \,\Big| \, 
    \frac{1}{\sqrt{1-\xi^2}}\, {\cal O}_{\bar{q}q}(\vec{0}) \Big|\,
              p^+, \vec{b}_0 \Big\rangle_{\sigma}
}
\nonumber \\
 &=& \int \frac{d^2\vec{p}'\, d^2\vec{p}}{(16\pi^3)^2}\;
   e^{i (\vec{p}'\cdot\vec{b}'_0 - 
                      \vec{p}\cdot\vec{b}_0^{\phantom{1}})}\;
   G\!\left(\vec{p}, \frac{1}{\sigma^2} \right)
   G\!\left(\vec{p'}, \frac{1}{\sigma^2} \right)\, 
                                {\cal H}(x,\xi,\vec{D}) ,
\end{eqnarray}
where for brevity we have omitted proton helicity labels.  Note that
at this point we need the definition of $\cal H$ in a frame with
arbitrary $\vec{p}$ and $\vec{p}'$.  Because $\cal H$ depends on these
vectors only through $\vec{D}$ we can carry out one of the
integrations after a change of variables, and find
\begin{eqnarray}
\lefteqn{
\rule[-1.3ex]{0pt}{1ex}_{\sigma} 
  \Big\langle p'^+, \vec{b}'_0 \,\Big| \, 
    \frac{1}{\sqrt{1-\xi^2}}\, {\cal O}_{\bar{q}q}(\vec{0}) \Big|\,
              p^+, \vec{b}_0 \Big\rangle_{\sigma}
 =  \frac{1}{16\pi^2 \sigma^2}\, \frac{(1-\xi^2)^2}{1+\xi^2}\,
     G\!\left( \delta\vec{b}, \half (1+\xi^2)\, \sigma^2 \right)\,
}
\nonumber \\[0.4em]
 && {}\times
  \int \frac{d^2\vec{D}}{16\pi^3}\; e^{-i\, \vec{D}\cdot\vec{b}} \,
  \exp\!\left[{-i}\, \frac{\xi(1-\xi^2)}{1+\xi^2}\,
                     \vec{D}\cdot\delta\vec{b} \right] 
  G\!\left( \vec{D}, \frac{2(1+\xi^2)}{(1-\xi^2)^2\, \sigma^2}
                 \right) \, {\cal H}(x,\xi,\vec{D})
\nonumber \\[0.4em]
 &\stackrel{\sigma\to 0}{=}& 
    \frac{1}{16\pi}\, (1-\xi^2)^2\; \deltwo(\delta\vec{b})\,
  \int \frac{d^2\vec{D}}{16\pi^3}\; e^{-i\, \vec{D}\cdot\vec{b}} \,
               {\cal H}(x,\xi,\vec{D}) ,
  \label{matrix-a}
\end{eqnarray}
where $\vec{b}$ and $\delta\vec{b}$ are given by
\begin{equation}
 \vec{b}_0  = -\frac{\vec{b}}{1+\xi} + \delta\vec{b} , \qquad
 \vec{b}'_0 = -\frac{\vec{b}}{1-\xi} - \delta\vec{b} .
   \label{positions}
\end{equation}
Up to smearing by $\delta\vec{b}$, constrained by a Gaussian to be of
order $\sigma$, the transverse positions of the the initial and final
state proton are not independent, and the crucial finding is that for
nonzero $\xi$ they are \emph{not the same}.  Technically this is due
to the fact that for given nonzero $\xi$ the matrix element $\cal H$
is not a function of $\vec{p}'-\vec{p}$ but of $\vec{D}$, which we saw
to be a direct consequence of Lorentz invariance.  More physically,
the transverse location of the proton changes when a finite
longitudinal momentum is transferred, which will become evident in
section~\ref{sec:overlap}.  Rewriting (\ref{matrix-a}) with
$\delta\vec{b}=0$ we obtain the main result of this section:
\begin{eqnarray}
\lefteqn{
\int \frac{d^2\vec{D}}{(2\pi)^2}\; e^{-i\, \vec{D}\cdot\vec{b}}\, 
    G(\vec{D},\, \vec{D}^2_\sigma)\, {\cal H}(x,\xi,\vec{D}) 
}
\nonumber \\
 &=& 64\pi^3 \sigma^2\,  \frac{1+\xi^2}{(1-\xi^2)^{5/2}}\:\;
   \rule[-1.3ex]{0pt}{1ex}_{\sigma} \Big\langle p'^+, 
     -\frac{\vec{b}}{1-\xi} \,
         \Big|\, {\cal O}_{\bar{q}q}(\vec{0}) \,\Big|\,
          p^+, -\frac{\vec{b}}{1+\xi} \Big\rangle_{\sigma}
\nonumber \\
 &=& 64\pi^3  \sigma^2\, \frac{1+\xi^2}{(1-\xi^2)^{5/2}}\:\;
   \rule[-1.3ex]{0pt}{1ex}_{\sigma} \Big\langle p'^+, 
     -\frac{\xi \vec{b}}{1-\xi} \,
        \Big|\, {\cal O}_{\bar{q}q}(\vec{b}) \,\Big|\,
         p^+, \frac{\xi \vec{b}}{1+\xi} \Big\rangle_{\sigma} \; ,
  \label{matrix-final}
\end{eqnarray}
where in the second step we have used translation invariance in the
transverse plane, and where
\begin{equation}
\sigma^2 = \frac{2(1+\xi^2)}{(1-\xi^2)^2}\, 
               \frac{1}{\vec{D}^2_\sigma} .
  \label{inverse}
\end{equation}
Coming back to the freedom of choice mentioned at the end of
section~\ref{sec:impact}, we remark that one could write
(\ref{matrix-a}) to (\ref{matrix-final}) in terms of a different
impact parameter variable, related to $\vec{b}$ by a $\xi$-dependent
scale factor.  This would of course not change the actual values of
the transverse distances described in these equations.


\subsection{Discussion}
\label{sec:discuss}

We now have an interpretation of the Fourier transformed GPDs
introduced in (\ref{I-def}) by taking the limit $\sigma\to 0$ of
(\ref{matrix-final}).  Of course its right-hand side does not vanish
in this limit, since the normalization (\ref{proton-norm}) of our
proton states includes a factor $1/\sigma^2$.  As data on GPDs will in
practice only be available in a finite range of $t$, one might
consider to actually perform the Fourier transform as given in
(\ref{matrix-final}), with $\vec{D}^2_\sigma$ of order of the largest
measured value $|t|_{\rm max}$ of $|t|$.  The Gaussian damping factor
would limit the effects of extrapolating the integrand to unmeasured
values of $\vec{D}^2$ or cutting off the integral.
Relations~(\ref{eq:t}) and (\ref{inverse}) give the accuracy $\sigma$
to which we can localize information in impact parameter space as
$\sigma \sim (|t|_{\rm max}- |t|_{\rm min})^{-1/2}$ with $|t|_{\rm
min} = 4\xi^2 m^2 /(1-\xi^2)$.

\begin{figure}
\begin{center}
  \leavevmode
  \epsfxsize=0.65\textwidth
  \epsfbox{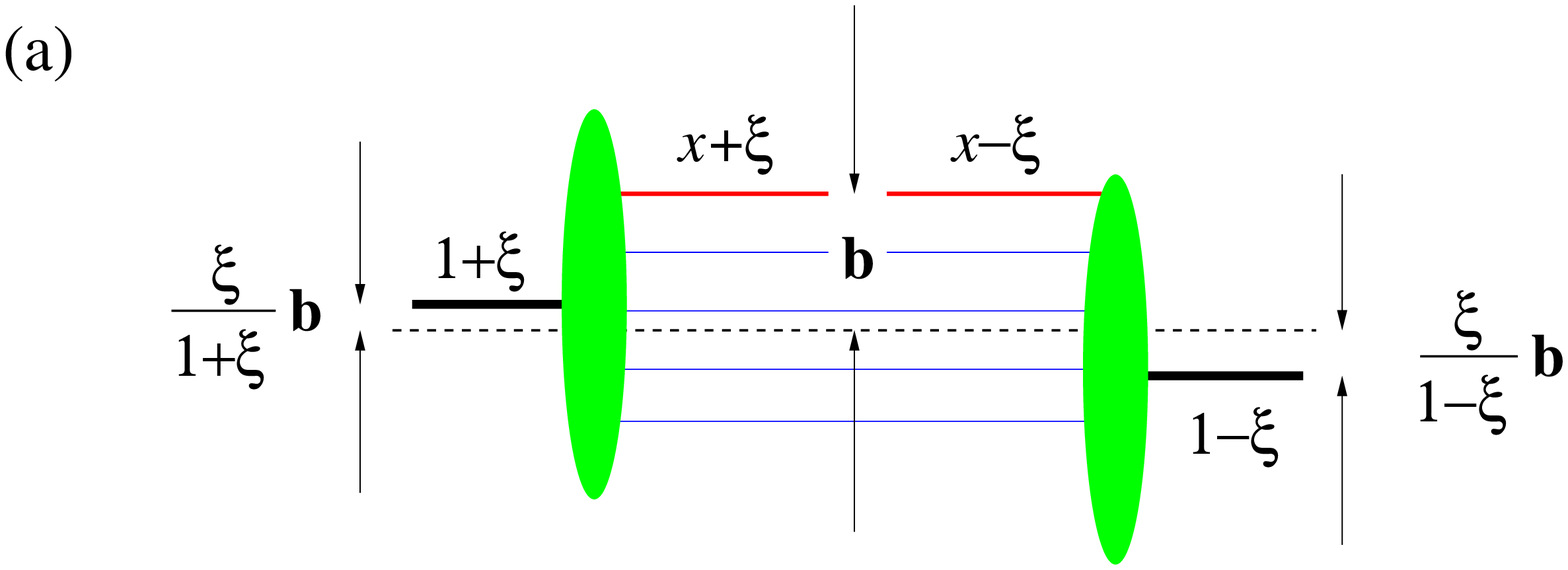}
\end{center}

\vspace{1em}

\begin{center}
  \leavevmode
  \epsfxsize=0.65\textwidth
  \epsfbox{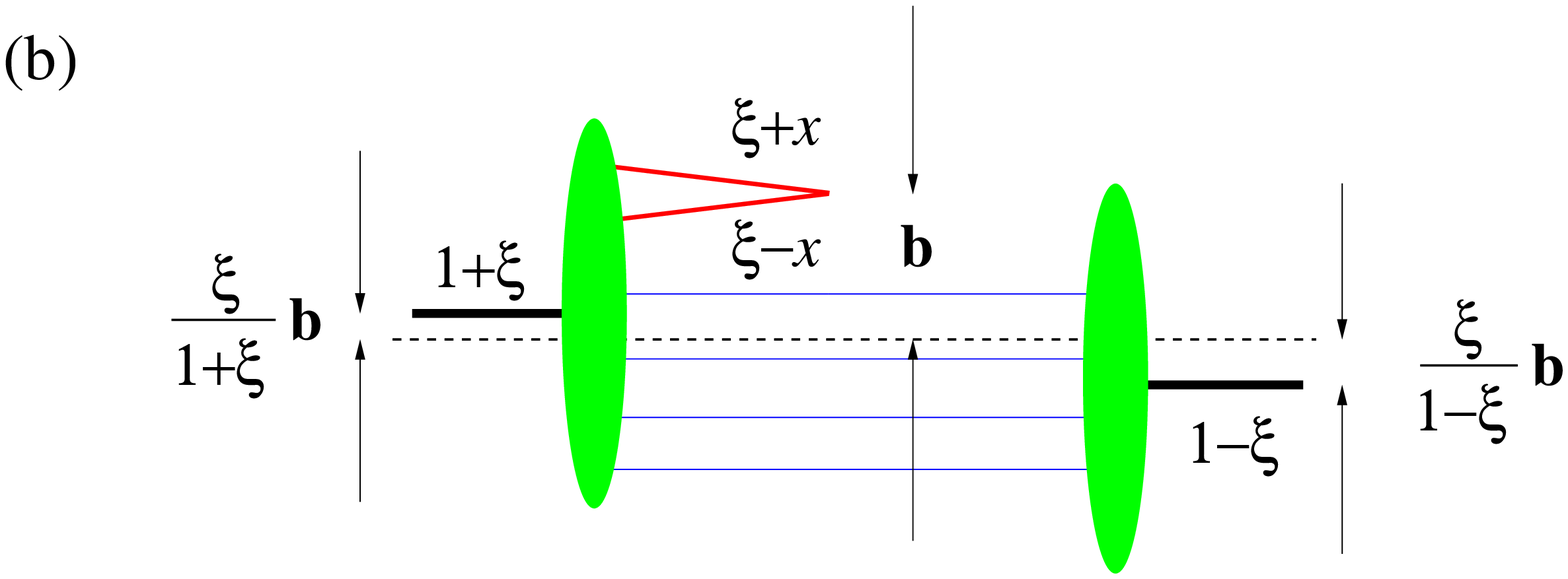}
\end{center}
\caption{\label{fig:impact} Representation of a GPD in impact
parameter space.  Plus-momentum fractions refer to the average proton
momentum $\frac{1}{2}(p+p')$ and are indicated above or below lines.
The region $\xi\le x\le 1$ is shown in (a), and the region
$|x|\le\xi$ in (b).}
\end{figure}

Figure~\ref{fig:impact} shows the physical picture encoded in
(\ref{matrix-final}).  GPDs in impact parameter space probe partons at
transverse position $\vec{b}$, with the initial and final state proton
localized around $\vec{0}$ but shifted relative to each other by an
amount of order $\xi\vec{b}$.  At the same time, the longitudinal
momenta of the protons and hadrons are specified, in the same way as
in the $t$-dependent GPDs.  In the DGLAP regions $x \in [\xi,1]$ and
$x \in [-1,-\xi]$, the impact parameter gives the location where a
quark or antiquark is pulled out of and put back into the proton.  In
the ERBL region $x \in [-\xi,\xi]$ the impact parameter describes the
transverse location of a quark-antiquark pair in the initial proton.

We remark that the shift of the transverse proton positions depends on
$\xi$ but not on $x$.  The information on the transverse location of
partons in the proton is therefore not ``washed out'' when GPDs are
integrated over $x$ with a weight depending on $x$ and~$\xi$.  As
anticipated in \cite{Ralston:2001xs} this implies that information is
already contained in the scattering amplitudes of hard processes,
where GPDs enter through just such a convolution in $x$.  Notice also
that for very small $\xi$ the difference between the proton positions
becomes negligible compared to the impact parameter $\vec{b}$ of the
struck quark.  In this sense the situation is simpler for the impact
parameter than for the longitudinal momentum fractions $x+\xi$ and
$x-\xi$.  Even at small $\xi$ their difference cannot be neglected
when $x$ is of order $\xi$, which it typically \emph{is} in the
convolution with the hard scattering kernel of a physical process.


\section{The overlap representation in $\protect\vec{b}$-space}
\label{sec:overlap}

A physical interpretation of momentum space GPDs has been obtained in
\cite{Diehl:2000xz,Brodsky:2000xy} by writing them in terms of
hadronic light-cone wave functions.  We will now derive the analog of
this in the mixed representation of definite plus-momentum and impact
parameter.

\subsection{Fock space and wave functions}
\label{sec:Fock}

A wave function representation of GPDs is naturally derived in the
framework of light-cone quantization and in the gauge $A^+=0$.  We
will only recall essentials here; more detail can for instance be
found in \cite{Diehl:2000xz} or in \cite{Brodsky:1989pv}.  In
light-cone quantization the dynamically independent components of a
quark field are given by the projection $q_+ = \frac{1}{2}
\gamma^-\gamma^+ q$.  At a given light-cone time, which we take as
$z^+=0$, this can be expanded in terms of creation and annihilation
operators for quarks and antiquarks of definite momentum,
\begin{eqnarray}
  \label{expand-mom}
q_+(0,z^-,\vec{z}) &=& \int \frac{d^2\vec{k}\, d k^+}{16\pi^3 k^+}\, 
  \theta(k^+)\, \sum_{\mu} \Big[\, 
      b(k^+, \vec{k}, \mu)\, u_+(k^+,\mu)\, 
           e^{-i k^+z^- + i\vec{k}\cdot\vec{z}}
\nonumber \\
 && \hspace{8.3em}
    {}+ d^\dag(k^+, \vec{k}, \mu)\, v_+(k^+,\mu)\, 
           e^{i k^+z^- - i\vec{k}\cdot\vec{z}} \,\Big] ,
\end{eqnarray}
where $\mu$ specifies the parton light-cone helicity.  Operators for
quarks and antiquarks with definite impact parameter are given by
\begin{eqnarray}
  \label{def-creator}
\tilde{b}(k^+, \vec{b}, \mu) &=& \int \frac{d^2\vec{k}}{16\pi^3}\,
    b(k^+, \vec{k}, \mu)\, e^{i\vec{k}\cdot\vec{b}} ,
\nonumber \\
\tilde{d}(k^+, \vec{b}, \mu) &=& \int \frac{d^2\vec{k}}{16\pi^3}\,
    d(k^+, \vec{k}, \mu)\, e^{i\vec{k}\cdot\vec{b}} .
\end{eqnarray}
The good components of the quark field can then be rewritten as 

\begin{eqnarray}
  \label{expand-imp}
q_+(0,z^-,\vec{b}) &=& \int \frac{dk^+}{k^+}\, 
  \theta(k^+)\, \sum_{\mu} \Big[\,
        \tilde{b}(k^+, \vec{b}, \mu)\, u_+(k^+,\mu)\, 
           e^{-i k^+z^-}
\nonumber \\
 && \hspace{6.5em}
    {}+ \tilde{d}^\dag(k^+, \vec{b}, \mu)\, v_+(k^+,\mu)\, 
           e^{i k^+z^-} \,\Big] .
\end{eqnarray}
Notice that for this to work it is essential that the good spinor
components $u_+(k^+,\mu) = \frac{1}{2} \gamma^-\gamma^+ u(k,\mu)$ and
$v_+(k^+,\mu) = \frac{1}{2} \gamma^-\gamma^+ v(k,\mu)$ are independent
of $\vec{k}$.  In other words, for the good field components one can
specify the light-cone helicity of a parton without specifying its
transverse momentum.  In full analogy to (\ref{def-creator}) one can
define annihilation operators $\tilde{a}(k^+, \vec{b}, \mu)$ for
gluons with definite impact parameter from their momentum space
counterparts $a(k^+, \vec{k}, \mu)$.

The Fock state expansion of a hadron state in momentum space reads
\begin{eqnarray}
  \label{Fock-mom}
\lefteqn{
|p^+, \vec{p}, \lambda\rangle =  \sum_{N, \beta}\, 
  \frac{1}{\sqrt{f_{N\beta}}} \int\prod_{i=1}^N\, 
  \frac{d x_i}{\sqrt{x_i}}\, \delta\Big( 1 - \sum_{i=1}^N x_i \Big)\;
  \frac{1}{(16\pi^3)^{N-1}}\, \prod_{i=1}^N d^2\vec{k}_i\; 
            \deltwo\Big( \vec{p} - \sum_{i=1}^N \vec{k}_i\Big )\;
}
\nonumber \\
&\times& \psi^\lambda_{N\beta}(x_i, \vec{k}_i - x_i \vec{p}) \,
  \prod_{i=1}^{N_q} \, b^\dag(x_i p^+, \vec{k}_i)
  \prod_{i=N_q+1}^{N_q+N_{\bar{q}}} 
                 \hspace{-0.8ex} d^{\,\dag}(x_i p^+, \vec{k}_i)
  \prod_{i=N_q+N_{\bar{q}}+1}^{N} 
                 \hspace{-1.2ex} a^\dag(x_i p^+, \vec{k}_i)
  \, |\, 0 \rangle 
\nonumber \\
\end{eqnarray}
using the conventions of \cite{Diehl:2000xz}.  $N$ specifies the
number of partons in a Fock state, and $\beta$ collectively labels its
parton composition and the discrete quantum numbers (flavor, color,
helicity) for each parton.  $f_{N\beta}$ is a normalization constant
providing a factor $n!$ for each subset of $n$ partons with identical
discrete quantum numbers; for ease of writing we have omitted labels
for these quantum numbers in the creation operators.  Notice that the
wave functions depend on the momentum variables of each parton only
via the light-cone momentum fraction $x^{\phantom{+}}_i = k_i^+ /p^+$
and its transverse momentum $\vec{k}_i - x_i \vec{p}$ relative to the
one of the hadron.\footnote{The momentum $\vec{k}_i - x_i \vec{p}$ is
obtained from $\vec{k}_i$ by a transverse boost (\ref{transv-boost})
to the frame where the parent hadron has no transverse momentum.}

Using the inverse transforms of (\ref{def-creator}) it is easy to
obtain from (\ref{Fock-mom}) the Fock state decomposition for a
transversely localized proton state (\ref{proton-sharp}) in terms of
transversely localized partons:
\begin{eqnarray}
  \label{Fock-imp}
\lefteqn{
|p^+, \vec{b}, \lambda\rangle =  \sum_{N, \beta}\, 
  \frac{1}{\sqrt{f_{N\beta}}} \int\prod_{i=1}^N\, 
  \frac{d x_i}{\sqrt{x_i}}\, \delta\Big( 1 - \sum_{i=1}^N x_i \Big)\;
  (4\pi)^{N-1}\, \prod_{i=1}^N d^2\vec{b}_i\;
            \deltwo\Big( \vec{b} - \sum_{i=1}^N x_i \vec{b}_i\Big )\;
}
\nonumber \\
&\times& \widetilde{\psi}^\lambda_{N\beta}(x_i, \vec{b}_i - \vec{b})\,
  \prod_{i=1}^{N_q} \, \tilde{b}^\dag(x_i p^+, \vec{b}_i)
  \prod_{i=N_q+1}^{N_q+N_{\bar{q}}} 
             \hspace{-0.8ex} \tilde{d}^{\,\dag}(x_i p^+, \vec{b}_i)
  \prod_{i=N_q+N_{\bar{q}}+1}^{N} 
             \hspace{-1.2ex} \tilde{a}^\dag(x_i p^+, \vec{b}_i)
  \, |\, 0 \rangle .
\nonumber \\
\end{eqnarray}
The wave functions for definite transverse momentum or impact
parameter are related by
\begin{eqnarray}
\widetilde\psi_{N\beta}^\lambda(x_i, \vec{b}_i - \vec{b})
  &=& \int [d^2 \vec{k}]_N\,
   \exp\!\Big[ i \sum_{i=1}^N \vec{k}_i\cdot\vec{b}_i \Big]\, 
  \psi^\lambda_{N\beta}(x_i, \vec{k}_i)
\hspace{3em} \mbox{with~} \vec{b} = \sum_{i=1}^N x_i \vec{b}_i \, ,
\nonumber \\[0.3em]
\psi_{N\beta}^\lambda(x_i, \vec{k}_i - x_i \vec{p})
  &=& \int [d^2 \vec{b}]_N\,
   \exp\!\Big[ {-i} \sum_{i=1}^N \vec{k}_i\cdot\vec{b}_i \Big]\, 
  \widetilde\psi^\lambda_{N\beta}(x_i, \vec{b}_i)
\hspace{1.8em} \mbox{with~} \vec{p} = \sum_{i=1}^N \vec{k}_i \, ,
\nonumber \\
\end{eqnarray}
and normalized to
\begin{equation}
P^\lambda_{N\beta}
\;=\; \int [d x]_N\, [d^2 \vec{k}_N]\;
  \Big| \psi^\lambda_{N\beta}(x_i, \vec{k}_i) \Big|^2
\;=\; \int [d x]_N\, [d^2 \vec{b}_N]\;
 \Big| \widetilde\psi^\lambda_{N\beta}(x_i, \vec{b}_i) \Big|^2 ,
\end{equation}
where $P^\lambda_{N\beta}$ is the probability to find the
corresponding Fock state in the proton, so that in total
$\sum_{N,\beta} P^\lambda_{N\beta} = 1$.  Here we have used the
shorthand notation
\begin{eqnarray}
[d^2 \vec{k}]_{N} &=& \frac{1}{(16\pi^3)^{N-1}} \prod_{i=1}^N 
   d^2\vec{k}_i\; \deltwo\Big( \sum_{i=1}^N \vec{k}_i \Big) ,
\nonumber \\{}
[d^2 \vec{b}]_{N} &=& (4\pi)^{N-1} \prod_{i=1}^N 
   d^2\vec{b}_i\; \deltwo\Big( \sum_{i=1}^N x_i \vec{b}_i \Big) ,
\nonumber \\{}
[d x]_N &=& 
   \prod_{i=1}^N d x_i\; \delta\Big( 1 - \sum_{i=1}^N x_i \Big) 
\end{eqnarray}
for the $N$-parton integration elements.

Let us briefly discuss the Fock state decomposition in impact
parameter form~(\ref{Fock-imp}).  The parton $i$ is transversely
localized at $\vec{b}_i$, but due to translation invariance the wave
function $\widetilde\psi_{N\beta}$ depends only on its position
relative to the center $\vec{b}$ of the proton.  The constraint
\begin{equation}
  \label{center-of-mom}
\vec{b} = \sum_{i=1}^N x_i \vec{b}_i
\end{equation}
identifies $\vec{b}$ as the ``center of plus-momentum''
\cite{Soper:1977jc} or ``transverse center of momentum''
\cite{Burkardt:2000wq} of the partons in each separate Fock state
$(N\beta)$.  This constraint is due to the invariance of the
light-cone formulation under transverse boosts.  Technically, it
arises in the derivation of (\ref{Fock-imp}) from (\ref{Fock-mom})
because the wave functions for proton states with different transverse
momenta $\vec{p}$ have the same functional form
$\psi^\lambda_{N\beta}(x_i, \vec{k}_i - x_i \vec{p})$.  Notice the
analogy of this situation with nonrelativistic mechanics
\cite{Soper:1977jc,Burkardt:2000wq}.  The transverse boosts
(\ref{transv-boost}) have the same form as a Galilean transformation
in two dimensions, with $k^+$ corresponding to the mass $m$ of a
particle , and $\vec{v}$ to the velocity characterizing the
transformation.  The conserved quantity corresponding to the
transverse center of momentum (\ref{center-of-mom}) is the center of
mass $\sum m_i \vec{r}_i /\sum m_i$ of an $N$-body system.


\subsection{The wave function representation of GPDs}
\label{sec:wave}

After the preparations of the preceding subsection it is easy to write
the matrix elements ${\cal H}_{\lambda'\lambda}$ and ${\cal
I}_{\lambda'\lambda}$ in terms of wave functions
$\widetilde\psi^\lambda_{N\beta}$.  This can be done either by
transforming the known representation of ${\cal H}_{\lambda'\lambda}$
in terms of $\psi^\lambda_{N\beta}$, or by repeating the derivation
given in \cite{Diehl:2000xz} directly in the impact parameter
formulation.  The result in the DGLAP region $x\in [\xi,1]$ is
\begin{eqnarray}
  \label{DGLAP-imp}
\lefteqn{
{\cal I}_{\lambda'\lambda}(x,\xi,\vec{b}) \;=\; \sum_{N,\beta}
  \sqrt{1-\xi^2}^{\, 1-N}\, \sum_{j=q}\;
  \int [d x]_N\, [d^2 \vec{b}]_N
} \hspace{5em}
\\[0.5em]
 &\times&  \delta(x-x_j)\; \deltwo\Big( \vec{b}-\vec{b}_j \Big)\;
  \widetilde\psi_{N\beta}^{*\, \lambda'}(x_i^{\rm out}, 
                \vec{b}_i^{\phantom{t}} - \vec{b}_0^{\rm out})\,
  \widetilde\psi_{N\beta}^\lambda(x_i^{\rm in}, 
                \vec{b}_i^{\phantom{t}} - \vec{b}_0^{\rm in})\,
\nonumber 
\end{eqnarray}
with wave function arguments
\begin{eqnarray}
  \label{fractions-DGLAP}
x_i^{\rm in}  &=&     \frac{x_i}{1+\xi} , \hspace{3.7em}
x_i^{\rm out} \;=\;   \frac{x_i}{1-\xi} 
                            \qquad\qquad \mbox{for~} i\neq j ,
\nonumber \\
x_j^{\rm in} &=&  \frac{x_j+\xi}{1+\xi} , \hspace{3.5em}
x_j^{\rm out} =\; \frac{x_j-\xi}{1-\xi} ,
\end{eqnarray}
and transverse locations of the proton states
\begin{eqnarray}
  \label{positions-DGLAP}
\vec{b}_0^{\rm in}  &=&    \frac{\xi}{1+\xi}\, \vec{b}_j , \qquad
\vec{b}_0^{\rm out} \;=\; -\frac{\xi}{1-\xi}\, \vec{b}_j . 
  \hspace{5.5em}
\end{eqnarray}
The label $j$ denotes the struck parton and is summed over all quarks
with appropriate flavor in a given Fock state, and the labels
$(N,\beta)$ are summed over the corresponding Fock states.  The
representation in the DGLAP region $x\in [-1,-\xi]$ is obtained
from~(\ref{DGLAP-imp}) by reversing the overall sign, by changing
$\delta(x-x_j)$ into $\delta(x+x_j)$, and by summing $j$ over
antiquarks.  In the ERBL region $x\in [-\xi,\xi]$ we have
\begin{eqnarray}
  \label{ERBL-imp}
\lefteqn{
{\cal I}_{\lambda'\lambda}(x,\xi,\vec{b}) = 
   \sum_{N,\beta,\beta'} 
   \sqrt{1-\xi}^{\, 2-N}\sqrt{1+\xi}^{\, -N}\, 
   \sum_{j,j'} \frac{1}{\sqrt{\rule{0pt}{1.4ex} n_j n_{j'}}}\, 
} \hspace{2em}
\\[0.2em]
 &\times& \int d x_j \prod_{i\neq j,j'}
     d x_i\; \delta\Big( 1- \xi - \sum_{i\neq j,j'} x_i \Big)
\nonumber \\[0.2em]
 &\times& (4\pi)^{N-1}\, 
  \int d^2\vec{b}_j \prod_{i\neq j,j'} d^2\vec{b}_i\;
     \deltwo\Big( \xi \vec{b}_j 
       + \sum_{i\neq j,j'} x_i \vec{b}_i \Big)
\nonumber \\[0.2em]
 &\times& \delta(x-x_j)\; \deltwo\Big( \vec{b}-\vec{b}_j \Big)\;
  \widetilde\psi_{N-1,\, \beta'}^{*\, \lambda'}(x_i^{\rm out}, 
                  \vec{b}_i^{\phantom{t}} - \vec{b}_0^{\rm out})\,
  \widetilde\psi_{N+1,\, \beta}^\lambda(x_i^{\rm in}, 
                  \vec{b}_i^{\phantom{t}} - \vec{b}_0^{\rm in})
\nonumber 
\end{eqnarray}
with $\vec{b}_0^{\rm in}$ and $\vec{b}_0^{\rm out}$ as in
(\ref{positions-DGLAP}) and 
\begin{eqnarray}
  \label{fractions-ERBL}
x_i^{\rm in}  &=&     \frac{x_i}{1+\xi} , \qquad
x_i^{\rm out} \;=\;   \frac{x_i}{1-\xi} 
                                \qquad \mbox{for~} i\neq j,j' ,
\nonumber \\
x_j^{\rm in}    &=&   \frac{\xi+x_j}{1+\xi} , \qquad 
x_{j'}^{\rm in} \;=\; \frac{\xi-x_j}{1+\xi} .
\end{eqnarray}
The partons $j,j'$ are the ones emitted from the initial proton, and
in (\ref{ERBL-imp}) one has to sum over all quarks $j$ and antiquarks
$j'$ with opposite helicities, opposite color, and appropriate flavor
in the initial state proton, over all Fock states $(N+1,\beta)$
containing such a $q\bar{q}$ pair, and over all Fock states $(N-1,
\beta')$ of the final state proton with matching quantum numbers for
the spectator partons $i\neq j,j'$.  The statistical factors $n_j$
($n_{j'}$) give the number of (anti)quarks in the Fock state
$(N+1,\beta)$ that have the same discrete quantum numbers as the
(anti)quark pulled out of the target.  The impact parameter
representations for quark helicity dependent and for gluon GPDs are
analogous to (\ref{DGLAP-imp}) and (\ref{ERBL-imp}); the relevant
modifications can easily be deduced from the momentum space
expressions in~\cite{Diehl:2000xz}.

If one defines the impact parameter GPDs with an additional Gaussian
weight $G(\vec{D},\, \vec{D}^2_\sigma)$ as in (\ref{matrix-final}) then
one simply has to replace 
\begin{equation}
\label{overlap-smeared}
 \deltwo\Big( \vec{b} - \vec{b}_j \Big) \to 
    \frac{{\vec{D}^2_\sigma}}{2\pi}\;
      G\!\left( \vec{b} - \vec{b}_j,\, 
                          \frac{1}{\vec{D}^2_\sigma} \right)
\end{equation}
in (\ref{DGLAP-imp}) and (\ref{ERBL-imp}).  This shows again that with
data for $\vec{D}^2$ up to order $\vec{D}^2_\sigma$ one can obtain
information in impact parameter space to an accuracy of order $1
/|\vec{D}_\sigma|$.  We remark that the precise implementation of this
is different in our result (\ref{matrix-final}), where the transverse
location of each proton is smeared out independently, and in
(\ref{overlap-smeared}), where the smearing is over the location of
the struck parton or parton pair, with a smearing of the
proton positions induced by (\ref{positions-DGLAP}).

The physical interpretation of the overlap formulae (\ref{DGLAP-imp})
and (\ref{ERBL-imp}) is the same as the one we have obtained in
section~\ref{sec:discuss} and illustrated in Fig.~\ref{fig:impact}.
We now see why for $\xi\neq 0$ the transverse location of the proton
is different before and after the scattering: according to our
discussion in subsection~\ref{sec:Fock}, its transverse center of
momentum is shifted because the proton is subject to a finite transfer
of plus-momentum.  We thus find that GPDs at nonzero $\xi$ correlate
hadronic wave functions with both different plus-momentum fractions
and different transverse positions of the partons.  Notice however
that the difference in transverse positions is a \emph{global} shift
in each wave function; the \emph{relative} transverse distances
between the partons in a hadron are the same before and after the
scattering.


\subsection{Positivity constraints}

Positivity constraints for GPDs in the impact parameter representation
have recently been considered by Pobylitsa in a very general framework
\cite{Pobylitsa:2002iu}.\footnote{Note that Pobylitsa defines impact
parameter GPDs as Fourier transforms with respect to the vector
$(1-\xi^2) \protect\vec{D}$, cf.~our remark at the end of
section~\protect\ref{sec:impact}.}
We shall not elaborate on this subject in detail, but only give a set
of inequalities that readily follow from the overlap representation
(\ref{DGLAP-imp}) for ${\cal I}_{\lambda'\lambda}$ in the DGLAP
region.  This representation has the structure of a scalar product
$(\phi_{-\xi}^{\,\lambda'} \,|\, \phi_{\xi}^{\lambda})$ in the Hilbert
space of wave functions, with
\begin{eqnarray}
\phi_{\xi}^{\lambda}(x_i,\vec{b}_i;N,\beta) &=& 
  \sqrt{1+\xi}^{\, 1-N}\,
  \widetilde\psi^\lambda_{N\beta}(x_i^{\rm in},
     \vec{b}_i - \vec{b}_0^{\rm in}) ,
\nonumber \\
\phi_{-\xi}^{\,\lambda'}(x_i,\vec{b}_i;N,\beta) &=& 
  \sqrt{1-\xi}^{\, 1-N}\,
  \widetilde\psi^{\lambda'}_{N\beta}(x_i^{\rm out},
     \vec{b}_i - \vec{b}_0^{\rm out}) .
\end{eqnarray}
For $\xi=0$ this implies that ${\cal I}_{++}(x,0,\vec{b})$ must be
real and positive if $x\ge 0$ and negative if $x\le0$, as observed in
\cite{Burkardt:2001ni}.  Furthermore we have the Schwartz inequality
$| (\phi_{-\xi}^{\,\lambda'} \,|\, \phi_{\xi}^{\lambda}) |^2 \le
(\phi_{-\xi}^{\,\lambda'} \,|\, \phi_{-\xi}^{\,\lambda'})\,
(\phi_{\xi}^{\lambda} \,|\, \phi_{\xi}^{\lambda})$, which after
suitable changes of integration variables in (\ref{DGLAP-imp}) gives
\begin{equation}
(1-\xi^2)^5\, \Big|\, 
  {\cal I}_{\lambda'\lambda}(x,\xi,\vec{b}) \,\Big|^2  \le 
{\cal I}_{++}\Big(\frac{x-\xi}{1-\xi},0, \frac{\vec{b}}{1-\xi}\Big)\;
{\cal I}_{++}\Big(\frac{x+\xi}{1+\xi},0, \frac{\vec{b}}{1+\xi}\Big)
\end{equation}
for $\xi\le x \le 1$ and any combination $\lambda',\lambda$ of proton
helicities.  Further relations are obtained by replacing ${\cal I}$
with the combinations ${\cal I} \pm \widetilde{\cal I}$ for definite
parton helicity.  More detailed inequalities involving the various
proton and parton helicity combinations can be obtained using the
matrix structure in the helicities, as shown in
\cite{Pobylitsa:2001nt}.


\section{Special cases}
\label{sec:special}

The wave function representation makes explicit in which way GPDs
probe the longitudinal and transverse structure of a hadron in a
correlated way.  In particular we have seen that for finite $\xi$ they
are correlations of wave functions where the partons differ both in
longitudinal momentum and in impact parameter by an amount controlled
by $\xi$.  In this section we briefly discuss the type of information
one obtains in some well-known special cases.

\subsection{The limit $\xi=0$}

In the case $\xi=0$ (where $\vec{D}$ coincides with the transverse
momentum transfer $\vec{p}'-\vec{p}$) the wave function representation
(\ref{DGLAP-imp}) becomes particularly simple, with
\begin{eqnarray}
  \label{forward-imp}
\lefteqn{
{\cal I}_{\lambda'\lambda}(x,0,\vec{b}) \,=\, \sum_{N,\beta}
  \sum_{j=q}\, \int [d x]_N\, [d^2 \vec{b}]_N\,
} \hspace{6em}
\\[0.5em]
 &\times&  \delta(x-x_j)\; \deltwo\Big( \vec{b}-\vec{b}_j \Big)\;
  \widetilde\psi_{N\beta}^{*\, \lambda'}(x_i, \vec{b}_i)\,
  \widetilde\psi_{N\beta}^\lambda(x_i, \vec{b}_i) .
\nonumber 
\end{eqnarray}
for $x\ge 0$ and an analogous relation for $x\le 0$. We see that if in
addition one takes the same spin state for the two proton states, the
Fourier transform of $H(x,0,t)$ is expressed through squared wave
functions and thus has a density interpretation.  For the Fourier
transform of $\tilde{H}(x,0,t)$ one obtains the difference of squared
wave functions for a right-handed and a left-handed struck quark or
antiquark. At $\xi=0$ the distribution $E$ only appears in the matrix
element for proton helicity flip, and its Fourier transform in ${\cal
I}_{-+}(x,0,\vec{b})$ correlates wave functions with identical parton
configurations but with opposite helicities of the parent proton.
Giving it a probability interpretation is more involved
\cite{Burkardt:2001ni}, and we will not elaborate on this issue.
Notice finally that no information is obtained about $\tilde E$ by
setting $\xi=0$ in the appropriate matrix elements, where due to its
definition (\ref{quark-pol-def}) it is accompanied by at least one
factor of~$\xi$.

Inverting the Fourier transform (\ref{I-def}) we see that the usual
quark density $q(x) = H(x,0,0)$ can be written as the integral of
${\cal I}_{++}(x,0,\vec{b})$ over $\vec{b}$, which removes the
$\deltwo( \vec{b}-\vec{b}_j )$ on the right-hand side of
(\ref{forward-imp}).  This makes it explicit that ordinary parton
densities specify the longitudinal momentum fraction of the struck
parton, averaged over its transverse position in the target.


\subsection{Form factors}

Integrating $H$, $E$, $\tilde{H}$, $\tilde E$ over $x$ one obtains the
elastic form factors of the quark vector and axial vector current.
Since the form factors are independent of $\xi$ one can evaluate their
wave function representation in a frame with $\xi=0$, which was the
original choice of Drell and Yan \cite{Drell:1970km}.  Introducing
Dirac and Pauli form factors for each separate quark flavor,
\begin{equation}
\langle p',\lambda'|\, \bar{q}(0) \gamma^\mu q(0) 
     \,|p,\lambda \rangle
= \bar{u}(p',\lambda') 
  \left[\, F_1^{q}(t)\, {\gamma}^{\mu} +
           F_2^{q}(t)\, \frac{i {\sigma}^{\mu\alpha}
                              (p'-p)_\alpha}{2m} \, \right]
  u(p,\lambda) ,
\end{equation}
we have impact parameter representations
\begin{eqnarray}
  \label{form-factors}
\lefteqn{
\frac{1}{4\pi} \int_{-\infty}^0 \! dt\, 
         J_0\Big( |\vec{b}| \sqrt{-t} \Big)\, F_1^{q}(t)
} \hspace{3em}
\\[0.3em]
 &=& \sum_{N,\beta}\, \sum_{j=q,\bar{q}} \sigma_j\, 
     \int [d x]_N\, [d^2 \vec{b}]_N\;
     \deltwo\Big( \vec{b}-\vec{b}_j \Big)\;
  \Big| \widetilde\psi_{N\beta}^{\lambda}(x_i, \vec{b}_i) \Big|^2 ,
\nonumber \\[0.5em]
\lefteqn{
   \frac{1}{4\pi}\, \frac{b^2 - i b^1}{|\vec{b}|}
         \int_{-\infty}^0 dt\,
         J_1\Big( |\vec{b}| \sqrt{-t} \Big)\; 
              \frac{\sqrt{-t}}{2m}\, F_2^{q}(t)
} \hspace{3em}
\nonumber \\[0.3em]
 &=& \sum_{N,\beta}\, \sum_{j=q,\bar{q}} \sigma_j\,
     \int [d x]_N\, [d^2 \vec{b}]_N\;
     \deltwo\Big( \vec{b}-\vec{b}_j \Big)\;
  \widetilde\psi_{N\beta}^{* \lambda'}(x_i, \vec{b}_i)\,
  \widetilde\psi_{N\beta}^{\lambda}(x_i, \vec{b}_i) ,
\nonumber 
\end{eqnarray}
where $\sigma_j$ is $+1$ for quarks and $-1$ for antiquarks, and
$\lambda=+\half$, $\lambda'=-\half$.  A corresponding relation exists
for the Fourier transform of the axial form factor.  The pseudoscalar
form factor decouples at $\xi=0$ and like $\tilde E$ requires
evaluation in a frame with nonzero~$\xi$.  We see that the Fourier
transform of the Dirac form factor $F_1^{q}$ has an immediate density
interpretation, whereas the Pauli form factor $F_2^{q}$ correlates
wave functions for opposite proton helicities.  To be precise, the
Fourier transforms in (\ref{form-factors}) describe the transverse
location of partons in a fast moving proton, irrespective of their
longitudinal momenta~\cite{Soper:1977jc}.  This is the opposite of
what we had for the usual parton densities, which contain purely
longitudinal information.  The interpretation of an elastic form
factor in a frame where the proton moves fast may seem unusual, but
note that we are describing hadron structure in terms of quarks,
antiquarks, and gluons here, so that such a frame is in fact very
adequate.

Taking higher moments in $x$ gives form factors of operators
containing derivatives.  In this case, information on longitudinal
structure is retained in the form of a momentum dependent weight.  The
simplest example is the quark part of the energy momentum tensor,
whose $(++)$ component can be written as \cite{Ji:1997ek}
\begin{eqnarray}
\langle p',\lambda'|\, T_q^{++}(0) \,|p,\lambda \rangle
 &=& \bar{u}(p',\lambda') 
  \left[\, A^{q}(t)\, P^{+} \gamma^{+}
         + B^{q}(t)\, \frac{P^{+} i{\sigma}^{+\alpha}
                      (p'-p)_\alpha}{2m} \right.
\nonumber \\
 & & \hspace{2.9em} \left. {}
         + C^{q}(t)\, \frac{(p'-p)^+ (p'-p)^+}{m}\, 
  \right] u(p,\lambda) 
\label{Ji12}
\end{eqnarray}
and is related with $\int dx\, x H$ and $\int dx\, x E$ by a sum rule.
In a frame with $\xi=0$ the form factor $C^q$ decouples and one is
left with $A^q$ and $B^q$.  Their representation in terms of momentum
space wave functions has been discussed in \cite{Brodsky:2000ii}.  In
impact parameter space it is obtained from (\ref{form-factors}) by
replacing $F_1^{q}$ with $A^q_{\phantom 1}$ and $F_2^{q}$ with
$B^q_{\phantom 1}$ on the left, and $\sigma_j$ with $x_j$ on the
right.  In other words, the contributions from quarks and antiquarks
$j$ now come with the same sign and weighted by their momentum
fractions.  In particular, the Fourier transform of $A^q$ describes
the transverse distribution of the longitudinal momentum carried by
quarks and antiquarks of a given flavor.  Integrating it over
$\vec{b}$ one obtains the second moment $\int dx\, x (q + \bar{q})$ of
quark and antiquark distributions.

\section{A tale of two scales}
\label{sec:scales}

So far we have suppressed the dependence of GPDs on the factorization
scale $\mu$, which in a physical process is provided by the hard scale
$Q$.  The dependence on this scale is given by well-known evolution
equations
\cite{Muller:1994fv,Ji:1997ek,Radyushkin:1997ki,Blumlein:1997pi} of
the general form
\begin{equation}
  \mu\, \frac{\partial}{\partial\mu} H(x,\xi,t;\mu) =
\int dy\; K(x,y,\xi;\alpha_s(\mu))\, H(y,\xi,t;\mu) ,
\end{equation}
with evolution kernels $K$ known up to two-loop
accuracy~\cite{Belitsky:1999hf}.  Important for us is that these
equations involve GPDs at the same $t$ and $\xi$, so that taking the
Fourier transform with respect to $\vec{D}$ does not alter their
structure.  The evolution equations for GPDs in impact parameter space
are hence the same as for their $t$-dependent counterparts.

The physical meaning of $\mu$ in our context is essentially as in the
case of ordinary parton distributions.  We recall that because of the
short-distance singularities of QCD, light-cone wave functions and the
underlying Fock state decomposition must be renormalized
\cite{Brodsky:1989pv}.  For our purpose it is useful to understand the
associated renormalization scale $\mu$ as a cutoff on transverse
momenta~\cite{Kogut:1974ub}.\footnote{Of course such a cutoff
regularization---which among other things breaks Lorentz
invariance---has its limitations, but it does make the physics
transparent.  There are other possible cutoff schemes, involving for
instance light-cone energy $k^-$ instead of transverse
momentum~\protect\cite{Brodsky:1989pv}. }
Quarks and gluons in the presence of this cutoff are then
``elementary'' down to a transverse resolution of order
$\mu^{-1}$---loosely speaking they have a transverse extension of that
size.  The wave functions in our overlap representations refer to
partons at a given scale $\mu$.  An impact parameter GPD in the DGLAP
region may thus be interpreted as describing the longitudinal momentum
and transverse location of a quark with transverse size~$\mu^{-1}$.
At larger values of $\mu$, this quark may be seen as consisting of a
quark plus a gluon, i.e., one will start to resolve its
``substructure''.  In the ERBL region a GPD in impact parameter space
describes a $q\bar{q}$-pair, where quark and antiquark are each of
size $\mu^{-1}$ and at the same transverse position in the proton, to
an accuracy again of order $\mu^{-1}$.

The role played by $\mu^{-1}$ is to be contrasted with the transverse
resolution $\sigma \sim (|t|_{\rm max}- |t|_{\rm min})^{-1/2}$
discussed in sections~\ref{sec:discuss} and \ref{sec:wave}, which
refers to the transverse location of a parton within its parent
hadron.  In other words, the momentum transfer $t$ is related to
\emph{where} a parton is found in the proton, whereas $\mu^2$
determines \emph{what} is meant by ``a parton''.  An ordinary quark
distribution at very large $\mu^2$, for instance, contains information
on the longitudinal momenta of quarks seen with very fine transverse
resolution, but no information at all on where quarks are located in
the transverse plane.  To use an analog from optics consider a cell
under a microscope.  Then $\mu^{-1}$ corresponds to the optical
resolution and limits which details of the cell one can see.  The
analog of $(|t|_{\rm max}- |t|_{\rm min})^{-1/2}$ specifies how
precisely one controls the position of the cell under the objective,
in order to determine where the magnified detail is located within the
cell.

At this point we can make a comment on the representation
(\ref{form-factors}) of the elastic form factors $F_1(t)$ and $F_2(t)$
in terms of wave functions for quarks, antiquarks and gluons.  At
small values of $-t$, say below 1~GeV$^2$, one may wonder how such a
description is possible, given that the form factors are measured in
elastic scattering processes whose momentum transfer $t$ is
insufficient to resolve any partonic structure at all.  The solution
of this apparent paradox is that these form factors are independent of
a renormalization scale $\mu^2$ because the quark vector current
$\bar{q}(z)\gamma^\mu q(z)$ is conserved.  They are hence the same
when evaluated at $\mu^2 \sim -t$ or at $\mu^2$ of several GeV$^2$.
Physically speaking, the transverse distribution of charge in the
proton is independent of how much substructure is resolved, so that
one may represent this charge distribution as due to quarks and
antiquarks, even though one has not resolved them explicitly.  In this
sense, $F_1(t)$ and $F_2(t)$ at small $t$ do contain information on
partons, but this information is not specific to these degrees of
freedom.

Notice that the situation is different for form factors of other
operators, which are given by higher moments of GPDs in $x$.  The
quark energy-momentum tensor for instance does depend on the
renormalization scale, and the information of its form factors is
specific to the value of $\mu$.  As $\mu$ increases, the average
momentum carried by quarks at a given impact parameter will become
smaller, since one resolves processes where the quarks lose momentum
by radiating gluons.  In the case of energy-momentum one still has a
conserved current when summing over all quark flavors and gluons, but
this does not hold for other local operators connected to moments of
GPDs.

\section{Unintegrated parton distributions}
\label{sec:unintegrated}

Apart from GPDs, there is another class of nonperturbative functions
that carry information not only on longitudinal but also on transverse
hadron structure.  These are $k_T$-dependent or unintegrated parton
distributions.  Let us see how the information they contain looks like
when expressed in terms of transverse position, and contrast it with
the picture we have obtained for GPDs.  For simplicity we will
restrict ourselves to forward distributions and set $p=p'$.  It is
then sufficient to consider a frame with $\vec{p}=0$.  Following the
naming scheme of~\cite{Mulders:1996dh} we define
\begin{equation}
  \label{quark-kt-def}
f_1(x,\vec{k}) =
 \int\frac{d^2\vec{z}\, d z^-}{16\pi^3}\;
 e^{i x p^{+}z^- - i \vec{k}\cdot \vec{z}}\;
\langle p,\lambda|
   \,\bar{q}(0,-\half z^-, -\half\vec{z})\,\gamma^+ 
         q(0,\half z^-,\half\vec{z})\, |p,\lambda\rangle ,
\end{equation}
which is related to the usual quark density by $\int d^2\vec{k}\,
f_1(x,\vec{k}) = q(x)$.  We suppress again the dependence on the
factorization scale $\mu$, whose discussion proceeds in analogy with
the preceding section.  A word of warning is in order concerning the
absence of the usual Wilson line between the operators $\bar{q}$ and
$q$.  These are now taken at different transverse positions $\pm
\half\vec{z}$, so that a Wilson line appears even in the gauge $A^+
=0$, except for particular choices of the integration path.  As
recently discussed in \cite{Collins:2002kn} there is subtle physics
encoded in the choice of path and the Wilson line.  This is beyond the
purpose of our present investigation, which is a basic understanding
of the spacial information contained in various types of parton
distributions.  The same holds for recent work where it was argued
that even for the usual parton distributions, with $\vec{z}=0$,
physical effects of the Wilson line are not correctly reproduced in
light-cone gauge~\cite{Brodsky:2002ue}.

\begin{figure}
\begin{center}
  \leavevmode
  \epsfxsize=0.65\textwidth
  \epsfbox{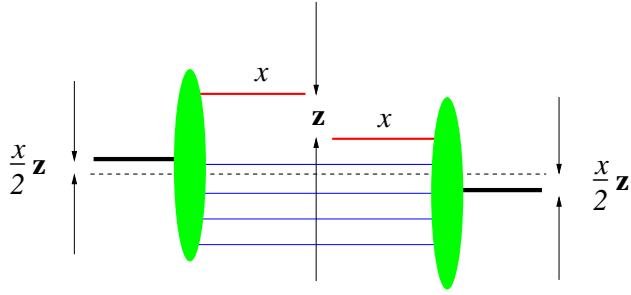}
\end{center}
\caption{\label{fig:k-perp} Impact parameter representation of an
unintegrated parton distribution.  $\vec{z}$~is the Fourier conjugate
variable to the transverse momentum $\vec{k}$ of the struck parton.}
\end{figure}

In transverse momentum space, the distribution $f_1(x,\vec{k})$ can be
interpreted as a probability density.  This is readily seen from its
wave function representation, which involves $| \psi(x_i, \vec{k}_i)
|^2$, with the wave function arguments of the struck parton fixed to
be $x_j=x$ and $\vec{k}_j= \vec{k}$.\footnote{In a frame with $\vec{p}
\neq 0$ one has instead $\vec{k}_j= \vec{k} - x \vec{p}$, in
accordance with the invariance of $\vec{k}_j$ under transverse boosts
(\protect\ref{transv-boost}).}
In terms of impact parameter wave functions we have
\begin{eqnarray}
  \label{k-perp-over}
\lefteqn{
\int d^2\vec{k}\; e^{i \vec{k}\cdot\vec{z}}\, f_1(x,\vec{k}) 
= \sum_{N,\beta}\, \sum_{j=q}\, \int [d x]_N\, [d^2 \vec{b}]_N
} \hspace{6em}
\\[0.5em]
 &\times&  \delta(x-x_j)\;
  \widetilde\psi_{N\beta}^{*\, \lambda}(x_i, 
                     \vec{b}^{\rm out}_i - \vec{b}_0^{\rm out})\,
  \widetilde\psi_{N\beta}^\lambda(x_i, 
                     \vec{b}^{\rm in}_i - \vec{b}_0^{\rm in})\,
\nonumber
\end{eqnarray}
for $x>0$, with wave function arguments
\begin{eqnarray}
\vec{b}_i^{\rm in} &=& \vec{b}_i^{\rm out} \;=\; 
                   \vec{b}_i  \hspace{3.5em} \mbox{~for~} i\neq j ,
\nonumber \\[0.3em]
\vec{b}_j^{\rm in}   &=&  \vec{b}_j + \frac{\vec{z}}{2} ,  \qquad
\vec{b}_j^{\rm out} \;=\; \vec{b}_j - \frac{\vec{z}}{2} , 
\end{eqnarray}
and proton states centered at
\begin{eqnarray}
\vec{b}_0^{\rm in}  &=&   x\, \frac{\vec{z}}{2} , \qquad \qquad
\vec{b}_0^{\rm out} \;=\; {}-x\, \frac{\vec{z}}{2} .
\end{eqnarray}
The corresponding physical picture is represented in
Fig.~\ref{fig:k-perp}.  As is already evident from the operator
defining $f_1(x,\vec{k})$ in (\ref{quark-kt-def}), we see that the
struck quark is at a relative transverse distance $\vec{z}$ in the
initial and the final state.  Unintegrated parton distributions thus
describe the correlation in transverse position of a single parton.
Notice that, in contrast to the situation for GPDs, the struck quark
now has a different transverse location \emph{relative} to the
spectator partons in the initial and the final state wave functions,
in addition to the overall shift of the proton center of momentum
dictated again by Lorentz invariance.  The dependence of the Fourier
transform (\ref{k-perp-over}) on $\vec{z}$ thus describes how much the
transverse location of a single parton can vary in the proton wave
function when all other partons are kept fixed.  Notice finally that
since $f_1(x,\vec{k})$ is measured at $t=0$, the overall position of
the struck quark with respect to the center of the proton is
integrated over in (\ref{k-perp-over}).

The quantities that contain this information in addition are of course
$k_T$ unintegrated GPDs.  They have been used in an estimate of power
corrections for hard exclusive processes in
\cite{Vanderhaeghen:1999xj} and recently been discussed in the context
of small-$x$ physics in~\cite{Martin:2001ms}.  Their impact parameter
representation combines the characteristic features of
(\ref{DGLAP-imp}) and (\ref{ERBL-imp}) with those of
(\ref{k-perp-over}).  In the DGLAP region, both the average transverse
position $\vec{b}$ and the relative offset $\vec{z}$ of the struck
parton in the initial and final state are now specified, whereas the
ERBL region describes a $q\bar{q}$-pair in the initial state proton,
with an average transverse position $\vec{b}$ and a relative
separation $\vec{z}$ between quark and antiquark.

\section{Conclusions}
\label{sec:sum}

A natural setting to represent the structure of a hadron in terms of
quarks and gluons is a frame where the hadron moves fast.  Its
direction of motion and the two perpendicular ones then acquire
different physical roles, and a comprehensive description of the
hadron requires information both on the longitudinal and the
transverse degrees of freedom.  Generalized parton distributions are
among the quantities that allow one to address such questions as:
``How broad is the spatial transverse distribution of fast quarks
compared with slow ones, and how does it compare with the spatial
transverse distribution of gluons?''

Whereas the description of scattering processes is in momentum space,
useful physical insight can be obtained in the mixed representation of
longitudinal momentum and transverse position, or impact parameter.
This representation is natural in various contexts, for instance in
the description of high-energy scattering processes, see
e.g.~\cite{Nachtmann:1996kt}, or the resummation of Sudakov logarithms
in hard reactions~\cite{Collins:1981uk}.  Burkardt has pointed out
that for zero skewness $\xi$, GPDs in impact parameter space have the
simple representation of a joint density in the longitudinal momentum
of a parton and its transverse distribution in the proton, at least
for helicity conserving transitions \cite{Burkardt:2000za}.  We find
that the situation for nonzero $\xi$ is similar up to an important
difference: as the proton loses longitudinal momentum its transverse
position is shifted by an amount proportional to $\xi$.  This is
because as a consequence of Lorentz invariance the ``transverse
position'' of the proton is the vector sum of the transverse positions
of its partons \emph{weighted} by their longitudinal momentum
fractions.  A more detailed picture is obtained by expressing GPDs in
terms of impact parameter wave functions, related to the ones in
momentum space by a Fourier transform.  In addition to their
longitudinal momentum fractions, the transverse distance of partons
from the proton center differs in the initial and final state, but
their relative distance to each other in a hadron stays the same.
This is in contrast to the case of $k_T$-dependent (but forward)
parton distributions, which in impact parameter space are correlation
functions for the transverse distance of a single parton with respect
to all other partons in the wave function.

All together, GPDs at nonzero $\xi$ describe quantum mechanical
correlations within a hadron rather than probability densities, both
in the transverse momentum and the impact parameter representations.
Despite their complexity in detail, the underlying physical picture is
simple, as shown in Fig.~\ref{fig:impact}.  As to achievable spatial
resolution, GPDs measured in a process with hard scale $Q^2$ provides
information on partons as seen with resolution $Q^{-1}$, whereas the
invariant momentum transfer $|t|$ to the proton must be known from its
minimum value $|t|_{\rm min}$ up to $|t|_{\rm max}$ in order to locate
partons in the transverse plane to an accuracy of order~$(|t|_{\rm
max}- |t|_{\rm min})^{-1/2}$.


\section*{Acknowledgments}

It is my pleasure to thank B.~Pire for discussions and R.~Jakob for
valuable remarks on the manuscript.



\newpage

\begin{center}
{\bf \Large Erratum}\\[2\baselineskip]

{\it Generalized parton distributions in impact parameter space
}\\[\baselineskip ]

Eur.\ Phys.\ J.\ {\bf C25}, 223 (2002), 
hep-ph/0205208 \\[0.5\baselineskip]

M.~Diehl \\[2\baselineskip]
\end{center}

The kinematical factor in the positivity bound (36) is incorrect.  The
bound correctly reads
\begin{displaymath}
(1-\xi^2)^3\, \Big|\, 
  {\cal I}_{\lambda'\lambda}(x,\xi,\vec{b}) \,\Big|^2  \le 
{\cal I}_{++}\Big(\frac{x-\xi}{1-\xi},0, \frac{\vec{b}}{1-\xi}\Big)\;
{\cal I}_{++}\Big(\frac{x+\xi}{1+\xi},0, \frac{\vec{b}}{1+\xi}\Big) .
\end{displaymath}
Our corrected result agrees with inequality (25) in
\cite{Pobylitsa:2002iu}, taking into account the different
normalization conventions here and there.


\begin{thebibliography}{10}

\bibitem{Muller:1994fv}
D.~M{\"u}ller, D.~Robaschik, B.~Geyer, F.~M. Dittes, and J.~Ho\v{r}ej\v{s}i,
\newblock Fortschr. Phys. {\bf 42}, 101 (1994), hep-ph/9812448.

\bibitem{Ji:1997ek}
X.-D. Ji,
\newblock Phys. Rev. Lett. {\bf 78}, 610 (1997), hep-ph/9603249.

\bibitem{Radyushkin:1997ki}
A.~V. Radyushkin,
\newblock Phys. Rev. {\bf D56}, 5524 (1997), hep-ph/9704207.

\bibitem{Burkardt:2000za}
M.~Burkardt,
\newblock Phys. Rev. {\bf D62}, 071503 (2000), hep-ph/0005108.

\bibitem{Ralston:2001xs}
J.~P. Ralston and B.~Pire,
\newblock Phys. Rev. {\bf D66}, 111501 (2002), hep-ph/0110075.

\bibitem{Diehl:2000xz}
M.~Diehl, T.~Feldmann, R.~Jakob, and P.~Kroll,
\newblock Nucl. Phys. {\bf B596}, 33 (2001),
Erratum-ibid.\ {\bf B605}, 647 (2001),
hep-ph/0009255.

\bibitem{Kogut:1970xa}
J.~B. Kogut and D.~E. Soper,
\newblock Phys. Rev. {\bf D1}, 2901 (1970).

\bibitem{Brodsky:1989pv}
S.~J. Brodsky and G.~P. Lepage,
\newblock {in: A.~H.~Mueller (Ed.),
\em {P}erturbative {Q}uantum {C}hromodynamics}, World
  Scientific, Singapore, 1989.

\bibitem{Diehl:2001pm}
M.~Diehl,
\newblock Eur. Phys. J. {\bf C19}, 485 (2001), hep-ph/0101335.

\bibitem{Brodsky:2000xy}
S.~J. Brodsky, M.~Diehl, and D.~S. Hwang,
\newblock Nucl. Phys. {\bf B596}, 99 (2001), hep-ph/0009254.

\bibitem{Soper:1977jc}
D.~E. Soper,
\newblock Phys. Rev. {\bf D15}, 1141 (1977).

\bibitem{Burkardt:2000wq}
M.~Burkardt,
\newblock hep-ph/0008051.

\bibitem{Pobylitsa:2002iu}
P.~V. Pobylitsa,
\newblock Phys. Rev. {\bf D66}, 094002 (2002), hep-ph/0204337.

\bibitem{Burkardt:2001ni}
M.~Burkardt,
\newblock hep-ph/0105324.

\bibitem{Pobylitsa:2001nt}
P.~V. Pobylitsa,
\newblock Phys. Rev. {\bf D65}, 077504 (2002), hep-ph/0112322; \\
%
P.~V. Pobylitsa,
\newblock Phys. Rev. {\bf D65}, 114015 (2002), hep-ph/0201030.

\bibitem{Drell:1970km}
S.~D. Drell and T.-M. Yan,
\newblock Phys. Rev. Lett. {\bf 24}, 181 (1970).

\bibitem{Brodsky:2000ii}
S.~J. Brodsky, D.~S. Hwang, B.-Q. Ma, and I.~Schmidt,
\newblock Nucl. Phys. {\bf B593}, 311 (2001), hep-th/0003082.

\bibitem{Blumlein:1997pi}
J.~Bl{\"u}mlein, B.~Geyer, and D.~Robaschik,
\newblock Phys. Lett. {\bf B406}, 161 (1997), hep-ph/9705264.

\bibitem{Belitsky:1999hf}
A.~V. Belitsky, A.~Freund, and D.~M{\"u}ller,
\newblock Nucl. Phys. {\bf B574}, 347 (2000), hep-ph/9912379.

\bibitem{Kogut:1974ub}
J.~B. Kogut and L.~Susskind,
\newblock Phys. Rev. {\bf D9}, 3391 (1974).

\bibitem{Mulders:1996dh}
P.~J. Mulders and R.~D. Tangerman,
\newblock Nucl. Phys. {\bf B461}, 197 (1996), hep-ph/9510301; \\
%
P.~J. Mulders,
\newblock hep-ph/9912493.

\bibitem{Collins:2002kn}
J.~C. Collins,
\newblock Phys. Lett. {\bf B536}, 43 (2002), hep-ph/0204004.

\bibitem{Brodsky:2002ue}
S.~J. Brodsky, P.~Hoyer, N.~Marchal, S.~Peign{\'e}, and F.~Sannino,
\newblock Phys. Rev. {\bf D65}, 114025 (2002), hep-ph/0104291.

\bibitem{Vanderhaeghen:1999xj}
M.~Vanderhaeghen, P.~A.~M. Guichon, and M.~Guidal,
\newblock Phys. Rev. {\bf D60}, 094017 (1999), hep-ph/9905372.

\bibitem{Martin:2001ms}
A.~D. Martin and M.~G. Ryskin,
\newblock Phys. Rev. {\bf D64}, 094017 (2001), hep-ph/0107149.

\bibitem{Nachtmann:1996kt}
O.~Nachtmann,
\newblock hep-ph/9609365; \\
%
M.~W{\"u}sthoff and A.~D. Martin,
\newblock J. Phys. {\bf G25}, R309 (1999), hep-ph/9909362.

\bibitem{Collins:1981uk}
J.~C. Collins and D.~E. Soper,
\newblock Nucl. Phys. {\bf B193}, 381 (1981),
Erratum-ibid.\ {\bf B213}, 545 (1983).

\end{thebibliography}
\end{document}